\documentclass[aps,letterpaper,twocolumn,preprintnumbers,floatfix,superscriptaddress]{revtex4}

\usepackage{simplewick}

\usepackage{graphicx}
\usepackage{dcolumn}
\usepackage{bm}
\usepackage{epsfig}
\usepackage{gensymb}
\usepackage{mathrsfs}
\usepackage{amsmath}
\usepackage{amssymb}
\usepackage{graphicx,bm}
\usepackage{slashed}
\usepackage{dsfont}
\usepackage[makeroom]{cancel}
\usepackage{youngtab}
\usepackage{multirow}
 \usepackage[titletoc]{appendix}

\def\fun#1#2{\lower3.6pt\vbox{\baselineskip0pt\lineskip.9pt
  \ialign{$\mathsurround=0pt#1\hfil##\hfil$\crcr#2\crcr\sim\crcr}}}

\def\lsim{\mathrel{\rlap{\raise 2.5pt \hbox{$<$}}\lower 2.5pt\hbox{$\sim$}}}
\def\gsim{\mathrel{\rlap{\raise 2.5pt \hbox{$>$}}\lower 2.5pt\hbox{$\sim$}}}

\input epsf

\newcommand{\be}{\begin{equation}}
\newcommand{\ee}{\end{equation}}
\newcommand{\bea}{\begin{eqnarray}}
\newcommand{\eea}{\end{eqnarray}}

\usepackage{color}

\newcommand{\comment}[1]{}

\begin{document}

\title{The Left-Right Symmetric Composite Higgs}

\author{Cong-Sen Guan}
\affiliation{CAS Key Laboratory of Theoretical Physics, Institute of Theoretical Physics,
Chinese Academy of Sciences, Beijing 100190, China.}
\affiliation{School of Physical Sciences, University of Chinese Academy of Sciences, Beijing 100190, P. R. China.}
\author{Teng Ma}
\affiliation{
Physics Department, Technion - Israel Institute of Technology, Haifa, 3200003, Israel}
\author{Jing Shu}
\affiliation{CAS Key Laboratory of Theoretical Physics, Institute of Theoretical Physics,
Chinese Academy of Sciences, Beijing 100190, China.}
\affiliation{School of Physical Sciences, University of Chinese Academy of Sciences, Beijing 100190, P. R. China.}
\affiliation{CAS Center for Excellence in Particle Physics, Beijing 100049, China.}
\affiliation{Center for High Energy Physics, Peking University, Beijing 100871, China}

\begin{abstract}
We find the exchange symmetry between left and right handed top quark in composite Higgs model with partial compositeness is efficient to soften the Higgs potential and reduce fine tuning.
This symmetry can keep the Higgs potential in top sector invariant under trigonometric parity $\sin (h/f) \leftrightarrow \cos (h/f)$ and thus the Higgs quadratic divergences can be completely cancelled, resulting a UV insensitive Higgs potential. We explicitly construct the minimal left-right symmetric model based on coset space $SO(6)/SO(5)$ which is locally isomorphic to $SU(4)/Sp(4)$ and thus has well defined fermionic UV completion. This UV completion can automatically keep Higgs potential in gauge sector finite even the gauge sector breaks this discrete symmetry. We find that the vector mesons can be very heavy while the colored top partners are relatively light ($>1.5$ TeV) to obtain a light Higgs.
\end{abstract}


\maketitle

\section{Introduction}
With the discovery of Higgs boson at LHC~\cite{Aad:2012tfa, Chatrchyan:2012xdj}, the Higgs mechanism~\cite{Englert:1964et, Higgs:1964pj, Higgs:1966ev,Guralnik:1964eu} in standard model (SM) is proven to successfully describe electroweak symmetry breaking (EWSB) at lower energy scale. Even though a great progress is achieved in understanding EWSB, some important issues are still mysterious for us. Among them the key mystery of EWSB is how to stabilize electroweak breaking scale against quantum correction.

In order to achieve successful EWSB, some new dynamics should be introduced at some high energy scale, such as TeV scale, to screen the quantum corrections from higher energy scales.  Among these new physics theories, the composite Higgs model (CHM)~\cite{Kaplan:1983fs,Georgi:1984af,Dugan:1984hq} in which Higgs is a pseudo-Nambu-Goldstone boson (pNGB) of some strong dynamics can solve it in a tremendously simple way.  Since Higgs is a bound state of strong dynamics, the quantum correction to Higgs potential from high energy scale is automatically screened by the constituents of Higgs, which results in the cut off of Higgs potential being reduced to the confinement scale $\Lambda$. Since pNGB Higgs potential is from loop corrections, if we further implement other mechanisms~\cite{ArkaniHamed:2001ca, Csaki:2017cep}, Higgs sensitivity to the confinement scale can be softened and Higgs can be light.

However, the successful EWSB requires tuning some parameters in the existing CHMs~\cite{Agashe:2004rs,DeCurtis:2011yx,Panico:2011pw,Marzocca:2012zn}. In these models, the main tuning is from separation between EWSB scale $v$ and global symmetry broken scale $f$ (called little hierarchy), $v \ll f$, in order to escape the constraints from electroweak precision tests~\cite{Peskin:1990zt,Barbieri:2004qk}. More worse, they always suffer from double tuning~\cite{Panico:2012uw}, which makes realistic Higgs potential very difficult to be achieved. Recently it is found that trigonometric parity~\cite{Csaki:2017jby} (TP) not only is the key ingredient to realizing neutral naturalness but also efficiently reduces the tuning in CHMs. The Higgs trigonometric parity can be induced not only through the exchange $Z_2$ symmetry between SM sector and hidden sector~\cite{Csaki:2017jby,Chacko:2005pe,TH2, Craig} but also through maximal symmetry~\cite{Csaki:2017cep,Csaki:2018zzf}.

In this paper, we find that the exchange symmetry between left-handed (LH) and right-handed (RH) SM fermions, called left-right (LR) symmetry, can eliminate the ultraviolet (UV) divergences and double tuning in the Higgs potential if partial compositeness~\cite{Kaplan:1991dc} is implemented. Different from the composite twin Higgs models, this model can realize TP in the Higgs potential without introducing any extra partners, but just through the left and right hand exchange symmetry of the SM fermions. As is well known, the TP is very efficient to eliminate UV divergence in the Higgs potential, this LR symmetry here enforces a UV insensitive Higgs potential in the top sector. The Higgs mass is correlated to the top partners mass, so usually 4 percent tuning is need to paid in order to get the light Higgs and heavy enough top partners in this model~\footnote{This minimal tunning can be reduced by raising the quartic term~\cite{Csaki:2019coc} or suppress the quadratic term~\cite{Csaki:2019qgb} in the Higgs potential.}. We find that the minimal coset to realize this LR symmetry is $SO(6)/SO(5)$~\footnote{Ref.~\cite{Marzocca:2012zn} chooses the Weinberg sum rule condition for CHMs based on $SO(5)/SO(4)$ that the Higgs UV divergence from the left-handed and right-handed fermions cancels. However, there is no real symmetry behind this Weinberg sum rule condition.} which is locally isomorphic to $SU(4)/Sp(4)$ and the latter has very well defined pure fermionic UV completions based on the confining hypercolor gauge theory.

The structure of this paper is as follows. In Sec. \ref{sec:LR_eff} we introduce the general method to realize LR symmetry from the perspective of low energy effective theory. In Sec. \ref{sec:eff} we explicitly realize this LR exchange symmetry in composite Higgs models based on the minimal coset $SO(6)/SO(5)$. In Sec. \ref{sec:SU4UV} we briefly discuss the $SU(4)/Sp(4)$ isomorphic version and the UV completion of this model. In Sec. \ref{sec:tuning} we analyse the EWSB and show that about 4 percent tuning is needed to achieve successful EWSB. Finally in Sec. \ref{sec:phenomonology} we briefly talk about the phenomenologies and showed that our model coincide with the direct search in experiment. The Appendix contain the generators, explicit form factors in effective Lagrangian and some comments on LR symmetric bottom sector.

\section{left-right symmetry in effective Lagrangian}
\label{sec:LR_eff}
In Twin Higgs models (THM)~\cite{Chacko:2005pe,TH2,Craig,Geller:2014kta,Low:2015nqa,Barbieri:2015lqa}, Higgs divergences are cancelled by introducing a $Z_2$ exchange symmetry between top and twin top to induce Higgs TP $\sin(h/f)\leftrightarrow\cos(h/f)$,  which is very efficient to cancel the quadratic divergences of the ordinary top. In this section, we show that this $Z_2$ parity can also be directly realized by an exchange symmetry between left and right handed top without introducing the twin sector. We first give a brief review to the trigonometric parity (TP) discussed in~\cite{Csaki:2017jby}. In pNGB Higgs models, any symmetric coset space $\mathcal{G}/\mathcal{H}$ contains a TP for any Goldstone boson $\pi^i$. This parity is the combination of $\pi/2$ rotation in $\pi^i$ direction with the Higgs parity ($\pi^i/f\rightarrow-\pi^i/f+\pi/2$) which leads to the transformation as
\bea
\cos \frac{\pi^i}{f} \leftrightarrow \sin \frac{\pi^i}{f} .
\eea
So if the left-right hand top exchange symmetry can contain the Higgs TP then the Higgs potential will be soften by this $Z_2$ symmetry, similar to THM. In table \ref{table:Z2_Symmetry}, we have listed different $Z_2$ symmetries used to induce Higgs TP in the maximal symmetry~\cite{Csaki:2017cep}, twin Higgs~\cite{Csaki:2017jby} and this left-right (LR) symmetric model.
\begin{table}[htp]
\begin{center}
\renewcommand\arraystretch{1.6}
\begin{tabular}{|c|c|c|c|}
  \hline
    & Maximal symmetry & Twin Higgs & LR model \\ \hline
  $Z_2$ & $t_{L,R}\leftrightarrow t_{L,R}$ & $t_{L,R}\leftrightarrow \tilde{t}_{L,R}$ & $t_L\leftrightarrow t_R$ \\ \hline
  Higgs TP & \multicolumn{3}{|c|}{$\sin(h/f)\leftrightarrow \cos(h/f)$} \\
  \hline
\end{tabular}
\end{center}
\caption{$Z_2$ symmetries to induce Higgs TP in different models.}
\label{table:Z2_Symmetry}
\end{table}
It's very easy to explicitly clarify the left-right symmetry and the mechanism to soft Higgs potential in low energy effective theory as we discussed as follows.


The minimal coset space to realize the left-right (LR) symmetry is $SO(6)/SO(5)$. In this model the electroweak (EW) gauge symmetry $SU(2)_L\times U(1)_Y$ is embedded in the $SO(4)$ which acts on the first four indices of $SO(6)$. In unitary gauge, only the physical Higgs boson $h$ is uneaten and the pNGB Higgs matrix $U$ is given by
\bea \label{eq:U}
U = \left(\begin{array}{cccc}
 \mathds{1}_3   &  &  &  \\
 & \cos \frac{h}{f}  &  & \sin \frac{h}{f} \\
 &  &  1 &  \\
 & -\sin \frac{h}{f} &  & \cos \frac{h}{f}
\end{array} \right),
\eea
where $f$ is the global symmetry broken scale. The Higgs TP operator of this model is~\cite{Csaki:2017jby}
\bea
P_1^h=R^h_{\pi/2} V= \left(\begin{array}{cccc}
 \mathds{1}_3   &  &  &  \\
 &   &  &-1\\
 &  &  1 &  \\
 & -1&  &
\end{array} \right),
\eea
where $R^h_{\pi/2}=U(h/f =\pi/2)$ is the $SO(2)$ rotation in Higgs direction with angle $\pi/2$ and $V=\mbox{diag}(1,1,1,1,1,-1)$ is the Higgs parity operator. It's easy to check $U$ transforms under $P_1^h$ as $U\rightarrow P_1^h U V=U(s_h\leftrightarrow c_h)$ with $s_h\equiv \sin(h/f)$ and $c_h\equiv\cos(h/f)$. In order to implement top LR symmetry, the EW doublet $q_L$ and singlet $t_R$ should be embedded in the same representation of $SO(6)$, such as fundamental representation:
\bea \label{eq:embedding}
\Psi_{q_L}&=&  \frac{1}{\sqrt{2}} \left( \begin{array}{c}
b_L \\
-i b_L \\
t_L \\
i t_L \\
0\\
0
\end{array} \right), \quad
\Psi_{t_{R} } =  \frac{1}{\sqrt{2}} \left( \begin{array}{c}
0 \\
0 \\
0 \\
0\\
t_R \\
 it_R
\end{array} \right),
\eea
where we choose the embeddings of the singlet $t_R$ has the similar form as $t_L$ (for the reason of constructing LR symmetry). Under this setup, the parity operator exchanging $t_L$ and $t_R$ should be
\begin{equation}\label{eq:P}
  P\equiv P_0P_1^h=\left(
                     \begin{array}{ccc}
                       \mathds{1}_2 &  &  \\
                        &  & \mathds{1}_2 \\
                        & \mathds{1}_2 &  \\
                     \end{array}
                   \right),
\end{equation}
where $P_0$ is operator that exchange the $3$rd and $5$th indices of $SO(6)$ and it's easy to check that $P_0$ acts trivially on $U$, $P_0UP_0=U$. The general effective Lagrangian invariant under global $SO(6)$ has the form
\begin{align}\label{eq:genLag}
  \mathcal{L}_\text{eff} & =\bar{\Psi}_{q_L}\slashed p(\Pi_0^q(p)+\Pi_1^q(p)\Sigma\Sigma^\dagger)\Psi_{q_L}\nonumber\\
   & +\bar{\Psi}_{t_R}\slashed p(\Pi_0^t(p) +\Pi_1 ^t(p) \Sigma \Sigma^\dagger) \Psi_{t_R} \nonumber \\
   & + M_1^t(p) \bar{\Psi}_{q_L} \Sigma \Sigma^\dagger \Psi_{t_R}+h.c.\, ,
\end{align}
where $\Sigma \equiv U.\mathcal{V} =(0, 0, 0, s_h, 0, c_h)^T$ is the linearly realized pNGB matrix which transforms under a global $SO(6)$ element $\bf g$ as $\Sigma \to {\bf g} \Sigma$ and $\mathcal{V} =(0,0,0,0,0,1)$ is the VEV breaking $SO(6)$ to $SO(5)$. From the effective Lagrangian we can formally inferred that in order to implement the LR exchange symmetry, the form factors in left and right handed top kinetic terms should be equal:
\bea \label{eq:identity}
\Pi_0 ^t(p) =\Pi_0 ^q(p), \quad \Pi_1 ^t(p) =\Pi_1 ^q(p),
\eea
which also implies that the underlying strong dynamics should be LR symmetric. Actually under the condition (\ref{eq:identity}), one can easily check that the effective Lagrangian is indeed invariant under the LR transformations defined as
\begin{align}\label{eq:LRTrans}
  \Psi_{q_L,t_R}&\rightarrow P\Psi_{t_R,q_L}=\Psi_{q_L,t_R}(t_L\leftrightarrow t_R),\nonumber\\
   \Sigma&\rightarrow P\Sigma=\Sigma(s_h\leftrightarrow c_h).
\end{align}
One can further find that if the form factors satisfy the relations in Eq.(\ref{eq:identity}), the kinetic terms in effective Lagrangian exhibits an enlarged global $SO(6) \times SU(6)$ symmetry where the QCD $SU(3)$ are embedded in $SU(6)$: $\bf 6 = 3 +3$ while the effective Yukawa coupling explicitly breaks it. We can write the effective Lagrangian in $SO(6) \times SU(6)$ form as
\bea \label{eq:effective_SU6}
\mathcal{L}_\text{eff} ^t&=& \text{Tr}[\bar{\Psi}_{LR} \slashed p(\Pi_0 (p) + \Pi_1 (p) \Sigma\Sigma^\dagger) \Psi_{LR}] \nonumber \\
&+& \text{Tr}[\bar{\Psi}_{LR}  M_1 ^t(p) \Sigma \Sigma^\dagger  \Psi_{RL}],
\eea
where $\Psi_{LR} =(\Psi_{q_L}, \Psi_{t_R})$ and  $\Psi_{RL} =(\Psi_{t_R},\Psi_{q_L} )$  are in ${\bf (6,6)}$ representation of  $\mathcal{G}' \equiv SO(6) \times SU(6)$ respectively. It is easy to find that the Yukawa coupling breaks this enlarged global symmetry. We can find these interactions are invariant under the LR $Z_2$ transformations in eq.(\ref{eq:LRTrans}) which can be rewritten as:
\bea
\Psi_{LR}  &\leftrightarrow&  P \Psi_{RL} = \Psi_{LR}(t_R \leftrightarrow t_L), \nonumber \\
 \Sigma &\to&  P\Sigma = \Sigma(s_h \leftrightarrow c_h).
\eea
So we have realized an exchange symmetry between $t_L$ and $t_R$ and this LR symmetry can trigger Higgs TP. The LR symmetry can be explicitly seen from the expansion of effective Lagrangian in terms of SM quarks:
\begin{align}
\label{eq:top_effective}
\mathcal{L}_\text{eff} &= \bar{b}_L \slashed p \Pi_0 ^q b_L +  \bar{t}_L \slashed p( \Pi_0 ^q + \frac{ \Pi_1 ^q}{2} s_h ^2 )t_L   \nonumber \\
 &+ \bar{t}_{R} \slashed p  ( \Pi_0 ^t + \frac{ \Pi_1 ^t}{2} c_h ^2 ) t_{R} +  \frac{ M_1 ^t }{2} \bar{t}_L t_R s_hc_h +h.c.
\end{align}
It's apparently invariant under $t_L\leftrightarrow t_R$, $s_h\leftrightarrow c_h$ when imposing (\ref{eq:identity}). Since the divergences of Higgs potential only from the Higgs dependant effective kinetic terms of top quark and there are only two pNGB fields ($\Sigma$) inserted in these terms (top quarks should be embedded in the simple representation of global symmetry), the quadratic divergent Higgs potential should be proportional to $s_h^2$ or $c_h^2$ and thus can  be cancelled by the Higgs TP induced from LR symmetry. We can directly calculate the one loop Higgs potential from top to prove our conclusion. In Euclidean momenta space, it is
\begin{align}\label{Vf}
&V_f = -2N_c \int \frac{d^ 4 p_E }{(2\pi)^4} \mbox{log}\left[1+ \frac{ \Pi_1 ^q(p_E)}{2 \Pi_0 ^q(p_E)} s_h ^2  + \frac{ \Pi_1 ^t(p_E)}{2 \Pi_0 ^t(p_E)} c_h ^2 \right. \nonumber \\
 &+ \left.  \frac{ \Pi_1 ^t(p_E) \Pi_1 ^q(p_E) }{4 \Pi_0 ^t(p_E) \Pi_0 ^q(p_E)}  s_h ^2c_h ^2 + \frac{ |M^t_1(p_E)|^2 }{4 p_E ^2 \Pi_0^t(p_E)  \Pi_0 ^q(p_E) } s_h^2 c_h ^2 \right],
\end{align}
where $N_c$ is QCD color number. If the top mass is produced through partial compositeness, at large Euclidean momenta region, $\Pi_0 ^{q,t} \propto p_E^0 $, $\Pi_1^{q,t} \propto p_E^{-2}$ and $M_1^t \propto p_E ^{-2}$ ~\cite{Marzocca:2012zn}. So the Higgs quadratic divergence in above Coleman-Weinberg potential according to power counting is
\bea
V_f \sim -2N_c \int \frac{d^ 4 p_E }{(2\pi)^4} \left[ \frac{ \Pi_1 ^q(p_E)}{2 \Pi_0 ^q(p_E)} s_h ^2  + \frac{ \Pi_1 ^t(p_E)}{2 \Pi_0 ^t(p_E)} c_h ^2 \right],
\eea
which indicates the Higgs quadratic divergence is only from top effective kinetic terms. Under the LR $Z_2$ symmetry, this term doesn't depend on Higgs field so Higgs quadratic divergence is eliminated.  Notice that the Higgs potential from top effective Yukawa coupling is always finite, which is the only term allowed by the maximal symmetry~\cite{Csaki:2017cep}.

In THM, the Higgs TP is induced by the exchange symmetry between top and its twin partner and Higgs potential is stabilized by the twin sector. While in LR symmetric CHM, the Higgs TP is induced by exchange symmetry between the LH and RH top and Higgs potential is stabilized by composite top partners.  So partial compositeness is necessary to realize natural Higgs potential for this model.

\section{The Minimal LR symmetric  Composite Higgs Model $SO(6)/SO(5)$}
\label{sec:eff}
According to the discussions in previous section, the minimal coset space to realize LR symmetry in top sector is $SO(6)/SO(5)$. However, the gauge sector doesn't exist LR symmetry so the Higgs potential from gauge bosons' loop is quadratic divergent without introducing new resonances. But since the coset space $SO(6)/SO(5)$ is locally isomorphic to the coset $SU(4)/Sp(4)$ which has very well defined fermionic UV completion, so the underlying strong dynamics can generally keep the Higgs potential  from gauge boson loops converge at high energy scale like pions in QCD. In order to mimic  this high energy behavior in the effective theory some heavy vector resonances should be introduced and impose Weinberg sum rules~\cite{Marzocca:2012zn} on these resonances' decay constants and spectrum.

Since the set up of gauge sector is the same as in paper~\cite{Csaki:2017jby}, we do not give any more discussions about it. To be self-contained  more details is shown in App.~\ref{app:gauge} (also can be found in~\cite{Csaki:2017jby}).

\subsection{Realization of LR symmetry}
In Sec.\ref{sec:LR_eff} we assumed certain properties (\ref{eq:identity}) in the low-energy effective Lagrangian and find LR symmetry can cancel Higgs quadratic divergence. In this section we will explain how to realize the LR symmetry in $SO(6)/SO(5)$ CHM with partial compositeness.

As discussed in Sec.\ref{sec:LR_eff}, that fermion masses are generated through partial compositeness is necessary for natural Higgs potential  so the composite top partners should be introduced. Since we have embedded LH doublet and RH singlet in the representation $\bf 6$ of $SO(6)$, the composite operator $\mathcal{O}$ that mix with top quark are also supposed to be the same representation in order to mix with top quarks. Since the composite sector is $SO(5)$ invariant so $\mathcal{O}$ should be decomposed to a five-plet and a singlet multiplets of $SO(5)$, $\mathcal{O}\rightarrow \Psi_\mathbf{5}+\Psi_\mathbf{1}$. The global $SO(6)$ invariant Lagrangian for elementary and composite fermions based on CCWZ  formalism~\cite{Wulzer:2016} can be constructed as
\bea
\label{eq:Lagrangian}
  \mathcal{L}_{f}&=&
  \bar{\Psi}_\mathbf{5}(i\slashed{\nabla}-M_5)\Psi_\mathbf{5}+\bar{\Psi}_\mathbf{1}(i\slashed{\nabla}-M_1)\Psi_\mathbf{1}\nonumber\\
  &+& f\bar{\Psi}_{q_L}P_{R}(\epsilon_{q1}U\Psi_\mathbf{1}+\epsilon_{q5}U\Psi_\mathbf{5})\nonumber\\
  &+&f \bar{\Psi}_{t_R}P_{L}(\epsilon_{t1}U\Psi_\mathbf{1}+\epsilon_{t5}U\Psi_\mathbf{5}) +h.c.\,,
\eea
where $\nabla_\mu = \partial_\mu -i E_\mu$ is covariant derivative of composite sector and $E_\mu$ is CCWZ $e$-symbol (we neglect the $U(1)_X$ charge $X$ with $Y = T_R ^3 +X$ in this model). The explicit embedding of composite resonance multiplets is
\bea
\label{eq:top_partners}
\Psi_\mathbf{5} =\frac{1}{\sqrt{2}}
\left(
\begin{array}{c}
 i B -iX_{5/3} \\
 B+X_{5/3} \\
 iT+iX_{2/3}\\
 -T+X_{2/3}\\
 i T_{-}^\prime-i T^\prime _+  \\
  0
\end{array}
\right),  \Psi_\mathbf{1} =\frac{1}{\sqrt{2}}
\left(
\begin{array}{c}
0 \\
 0\\
 0\\
 0\\
 0  \\
 T_{-}^\prime+T^\prime _+
\end{array}
\right)
\eea
where the two EW doublets with hypercharge $1/6$ and $7/6$,  $(T,B)$ and $(X_{5/3}, X_{2/3})$, fill the full representation of custodial symmetry $SO(4)$ and $T_{+,-} ^\prime$ are EW singlet with positive and negative $SO(2)_\eta$ charge respectively.

Before discussing the LR symmetry, we first need to figure out how the pNGB fields $U$ transform under the LR parity $P=P_0P_1^h$ we defined in Sec.\ref{sec:LR_eff}. Recall that $U$ transforms under the TP as ($U$ in unitary gauge) $P_1^h U V=U(s_h\leftrightarrow c_h)$ and $P_0$ acts trivially on $U$, so it's easy to derive that ($P_0^2=1$)
\bea
\label{eq:LR_symmetry}
P U =P_0 P_1 ^h U = U(s_h \leftrightarrow c_h) P_0V.
\eea
Use this identity, we find if and only if the linear mixing couplings of LH and RH top to the composite multiplets are correspondingly equal,
\bea
\label{eq:couplings}
\epsilon_{q1} =\epsilon_{t1} ,\quad  \epsilon_{q5} =\epsilon_{t5},
\eea
the Lagrangian is invariant under the LR exchange symmetry defined as
\begin{align}\label{eq:LRsymmetry}
  U & \rightarrow PUVP_0=U(s_h\leftrightarrow c_h) \nonumber\\
  \Psi_{q_L} & \leftrightarrow P\Psi_{t_R},\: P_R\Psi_{\mathbf{1},\mathbf{5}}\leftrightarrow P_0VP_L\Psi_{\mathbf{1},\mathbf{5}}.
\end{align}

Use the explicit matrix form of these parity operators, one can easily check that $P_0V$ is an element of the unbroken subgroup $SO(5)$, which means when it acts on the composite fermion multiplets, the effect is just a redefinition of each components (i.e. $\Psi_{\mathbf{1},\mathbf{5}}\rightarrow \Psi_{\mathbf{1},\mathbf{5}}^\prime$). So the LR exchange symmetry is a good symmetry of interactions between fermion and NGBs in Eq.(\ref{eq:Lagrangian}) and it will result in the Higgs potential from these fermions' loops is exactly TP invariant. After integrating out the composite fermions, we can get the LR symmetric effective Lagrangian (\ref{eq:genLag}) satisfying Eq.(\ref{eq:identity}) and the explicit expressions of the general form factors is shown in Appendix. \ref{App:generators}.

In Sec.\ref{sec:LR_eff} we demonstrated that the LR symmetry can cancel the quadratic divergence in Higgs potential based on the effective Lagrangian and partial compositeness. Here we can also analyse the Higgs potential directly from the composite Lagrangian (\ref{eq:Lagrangian}). Notice that since the SM fermions are embedded in the fundamental representation of global symmetry one $U$ field is always associated with one Yukawa coupling $\lambda \in f\{ \epsilon_{q1}, \epsilon_{q5} ,\epsilon_{t1}, \epsilon_{t5} \} $. According to power counting, the quadratic divergent terms are proportional to $\lambda^2$ so the quadratic divergent terms of Higgs potential from fermion sector at most contain two powers of $U$, proportional to $s_h ^2$ or $c_h ^2$. So Higgs TP is enough to eliminate them. However the log divergent terms can at most contain four powers of $U$, proportional to the fourth power of trigonometric function, so usually they can't be eliminated by TP. To further eliminate these terms, one can, as did in~\cite{Csaki:2017cep}, suppose the composite sector has fully $SO(6)$ global symmetry in Yukawa couplings and the composite mass terms explicitly break it to $H \equiv SO(5)$ or equivalently embed this model in the $SO(6)/SO(5)$ two-site moose, which requires the mixing couplings satisfy
\bea
\label{eq:fullso6}
\epsilon_{q1} =\epsilon_{q5}, \quad  \epsilon_{t1} =\epsilon_{t5}.
\eea
So $\Psi_\mathbf{5}$ and $\Psi_\mathbf{1}$ can be combined to fill complete representation \textbf{6} of global $SO(6)$. The Lagrangian with LR symmetry and fully global $SO(6)$ invariant elementary-composite mixing terms (two-site moose De-construction) thus has the form as follows
\begin{align}
  \mathcal{L}_f & = \bar{\Psi}_\mathbf{6}i \slashed{\nabla}\Psi_\mathbf{6}-\frac{M_5+M_1}{2}\bar{\Psi}_\mathbf{6}\Psi_\mathbf{6}-\frac{M_5
  -M_1}{2}\bar{\Psi}_\mathbf{6}V\Psi_\mathbf{6} \nonumber\\
   & +f\epsilon\bar{\Psi}_{q_L}P_RU\Psi_\mathbf{6}+f\epsilon\bar{\Psi}_{t_R}P_LU\Psi_\mathbf{6}+h.c.
\end{align}
Under this settings, the Higgs shift symmetry is only broken by the top partner's masses so the leading Higgs potential should be proportional to mass square $M^2$ with $M \in \{M_5, M_1 \}$. On the other hand, TP invariant Higgs potential at least contain four powers of $U$ thus is proportional to $f^4\epsilon^4$. Combining above analysis, we find that if the fermion-Higgs interactions are LR symmetric and the elementary-composite mixing terms are fully $SO(6)$ invariant (i.e. the Yukawa couplings satisfy both Eq.(\ref{eq:couplings}) and Eq.(\ref{eq:fullso6})) or equivalently is de-constructed on two-site moose, Higgs potential must be proportional to $f^4\epsilon^4 M^2$ thus finite according to power counting.

Above discussion is very general and doesn't depend on the particular coset space. Below we summarize the general principle to construct a TP invariant Higgs potential through LR exchange symmetry. Higgs TP $P_1^h$ always exists in any symmetric coset space $G/H$. To implement the LR symmetry and preserve Higgs TP, the LH and RH top should be embedded in the same representation of $G$ in such a way: the L-R exchange operator $P$ can be written as $P=P_0P_1^h$ containing Higgs TP parity operator. The LR symmetry requires the Yukawa couplings of LH and RH top to the composite resonances should be equal (i.e. chirality independent), which will make the form factors of LH and RH top kinetic terms equal in the lower-energy effective lagrangian. So this LR symmetry can preserve Higgs TP and can efficiently forbid $s_h ^2$ and $c_h ^2$ terms (actually $\sin^2(2N +1) h/f $ and $\cos^2(2N+1)h/f $ terms are all be forbidden, where $N$ is integer). If the quadratic divergent terms in Higgs potential are only proportional to these terms, TP can eliminate it. While to further make Higgs potential finite, the elementary-composite mixing terms should be fully $G$ invariant in the composite side or de-construct this model in two-site moose.

\section{UV completion for $SU(4)/Sp(4)$}\label{sec:SU4UV}
As is well known, $SO(6)/SO(5)$ and $SU(4)/Sp(4)$ are locally isomorphic and $SU(4)/Sp(4)$ has well defined UV completion based on a purely fermionic hypercolor theory. Since the Higgs sector is the same as the model in~\cite{Csaki:2017jby}, we just review main results in this sector and then focus on the UV completion for LR symmetric partial compositeness sector.
\subsection{Some reviews on UV completion for Goldstone sector}
The $SU(4)/Sp(4)$ breaking pattern can be realized by introducing four Weyl fermions $\psi_i$ with $i=1, 2, 3, 4$. The hypercolor gauge group can be chosen as $Sp(2N)$ and these preons are in the fundamental representation of hypercolor group~\cite{Galloway:2010bp, Cacciapaglia:2014uja}. The first two fermions $(\psi_1, \psi_2)$ are arranged into the $SU(2)_L$ doublet with $U(1)_\eta \cong SO(2)_\eta$ charge $\bf1$ and the last two fermions $(\psi_3, \psi_4)$ are $SU(2)_R$ doublet with  hypercharge $\bf \mp1/2$ respectively and $U(1)_\eta$ charge $\bf -1$. Since the preons are in pseudo-real representation, their scalar condensate $\langle \psi_i \psi_j \rangle \ne 0$ is antisymmetric with respect to $SU(4)$ flavor symmetry and thus it can break global symmetry $SU(4)$ into $Sp(4)$. This broken pattern produces five NGBs and their quantum number is same as the ones in $SO(6)/SO(5)$ under the electroweak unbroken vacuum. We choose this vacuum in the form of
\bea\label{eq:vacuum}
 V = \left( \begin{array}{cc}
i \sigma_2 & 0\\
0& -i \sigma_2 \\
\end{array}  \right).
\eea
In this vacuum, the automorphism map can be constructed as
\bea
T^a \to -V T^{aT} V^T,  \quad  T^{\hat{a}}  \to  -V T^{\hat{a} T} V^T,
\eea
where $T^a$ and $T^{\hat{a}}$ are unbroken and broken generators. With this map, the linearly realized Goldstone field $\Sigma^\prime$ can be constructed as
\bea
\Sigma^\prime = U^2V,
\eea
where $U$ is Goldstone matrix. It transforms linearly under global symmetry as $
\Sigma^\prime \to  g\Sigma^\prime g^T$ with $g \in SU(4)$.
\subsection{LR symmetric top sector in $SU(4)/Sp(4)$}
Since the $SU(4)/Sp(4)$ is locally isomorphic with $SO(6)/SO(5)$ and has UV completion of pure fermions and gauge bosons, we can translating the LR symmetric top sector into this coset to find the realization of LR symmetry in the UV completion.  According to the correspondence between this two equivalent coset spaces, the LH top doublet and RH top are embedded in two indices anti-symmetric representation $\bf 6$ of $SU(4)$:
\bea \label{eq:SU4_embedding}
\Psi_{q_L} = \frac{1}{\sqrt{2}} \left( \begin{array}{cc}
 {\bf 0} &  Q \\
-Q^T & \bf{ 0}  \\
\end{array}  \right),  \;  \Psi_{t_R}   =  \frac{1}{\sqrt{2}} \left( \begin{array}{cc}
 i t_R \sigma^2 &  0 \\
0 & \bf{ 0}  \\
\end{array}  \right),
\eea
where all of these are four by four antisymmetric matrices and $Q =(q_L,{\bf 0}_{2\times 1})$. The LR exchange symmetry requires the global symmetry in top sector is enlarged to $SU(4) \times SU(6)$, as the $SO(6)/SO(5)$ case. Based on this enlarged global symmetry, we can write the effective Lagrangian to explicitly show the LR symmetry:
\begin{align} \label{eq:effective_SU4}
\mathcal{L}_\text{eff} &=\Pi_0 (p) \mbox{Tr}[ \bar{\Psi}_{LR} \slashed p  \Psi_{LR} ] +\Pi_1(p) \mbox{Tr}[  \bar{\Psi}_{LR} \Sigma^\prime ] \slashed p \mbox{Tr}[ \Psi_{LR} \Sigma^{\prime \dagger}]      \nonumber \\
&+ M_1 ^t(p) \mbox{Tr}[ \bar{\Psi}_{LR}   \Sigma^\prime ]  \mbox{Tr}[ \Psi_{RL}   \Sigma^{\prime \dagger} ]
\end{align}
where  $\Psi_{LR} =(\Psi_{q_L} ,\Psi_{t_R} )$ and $\Psi_{RL} =( \Psi_{t_R},\Psi_{q_L})$ are in the ${\bf (6,6)}$ representations of $SU(4) \times SU(6)$, and again $\Pi_{0,1}$ and $M_{1}^t$ are form factors. The effective interactions are invariant under the LR exchange symmetry:
\begin{align} \label{eq:Z2_SU4}
\Sigma' &\to  P_1 \Sigma'  P_1 =\Sigma' (s_h \leftrightarrow c_h) \nonumber \\
\Psi_{LR} &\to   P_1  \Psi_{RL} P_1  =  \Psi_{LR}(t_L \leftrightarrow t_R) \nonumber \\
\Psi_{RL} &\to   P_1  \Psi_{LR} P_1  =  \Psi_{RL}(t_L \leftrightarrow t_R),
\end{align}
where the LR exchange operator $P_1$ is
\bea
P_1 &=& \left( \begin{array}{cccc}
1\\
&&1&\\
& 1& &  \\
 &  & &-1
\end{array}  \right),
\eea
which is an element of $SU(4)$. We can explicitly  express the effective Lagrangian in terms of SM quarks as
\bea \label{eq:top_effective2}
\mathcal{L}_\text{eff} ^t&=&\bar{b}_L\slashed p\Pi_0(p)b_L+\bar{t}_L\slashed p(\Pi_0(p)+2\Pi_1(p)s_h^2)t_L\nonumber\\
&+& \bar{t}_R \slashed p( \Pi_0(p)+2\Pi_1(p) c_h ^2 )t_R  \nonumber \\
&+& 2 M_1 ^t(p) \bar{t}_L t_R  s_h c_h   +h.c.
\eea
It is easy to see that this Lagrangian is LR symmetric and is equivalent to the effective Lagrangian in the $SO(6)/SO(5)$ case in Eq.(\ref{eq:top_effective}) up to a redefinition of the form factors.

\subsection{UV completion for partial compositeness with LR exchange symmetry}
In the previous two subsections, we have roughly argued the UV completion for gauge-Goldstone sector and realization of LR exchange symmetry in low-energy effective Lagrangian for top sector. In this section we will show the UV completion for effective top Yukawa. We suppose the effective Lagrangian arises from integrating out composite fermions which linearly mix with elementary top. To produce these bound states, some colored preons should be introduced. The colored preons can be either fermions or scalars. But we find scalar preons can not realize LR symmetry in fermion sector. So in the following, we will explain pure fermionic UV completion for partial compositeness.

As in~\cite{Csaki:2017jby}, in order to form the fermionic bound states with preons $\psi_i$, the colored Weyl fermions $\chi_{L,R}$ in two indices anti-symmetric representation of the $Sp(2N)$ hypercolor gauge group should be introduced. Their quantum numbers under hypercolor and SM gauge symmetry $Sp(2N) \times SU(3)_c  \times SU(2)_L \times U(1)_Y $ is summarized in Tab.~\ref{table:All_Quantum_Number}.
\begin{table}[htp]
\begin{center}
\begin{tabular}{c||cccc}
   & $Sp(2N)$ & $SU(2)_L$ & $U(1)_Y$& $SU(3)_c$  \\
     \hline \\
 $\chi_L$ & $\tiny{\yng(1,1)}$  & ${\bf 1}$& $\frac{2}{3}$ & ${\tiny{\yng(1)}}$ \\
 &&&& \\
 $\chi_R^c$ & $\tiny{\yng(1,1)}$  & ${\bf 1}$& $-\frac{2}{3}$ & ${\tiny{\bar{ \yng(1) }}}$  \\
  \\  \hline
\end{tabular}
\end{center}
\caption{The quantum numbers of the colored preons.}
\label{table:All_Quantum_Number}
\end{table}
According to the quantum number of the preons, the wave functions for top partners $\Psi_{\mathbf{5},\mathbf{1}}$ are
\bea
\{\Psi_{\mathbf{5} L}, \Psi_{\mathbf{1}L} \} \cong \chi_L (\psi \psi), \quad \{\Psi_{\mathbf{5} R}, \Psi_{\mathbf{1}R} \} \cong \chi_R (\psi \psi),
\eea
where the bracket $"()"$ denotes the Lorentz index contract of the fermions. As discussed in $SO(6)/SO(5)$ model, the LR symmetry in the low-energy effective top couplings requires the composite partners sector should be LR exchange invariant, see Eq.(\ref{eq:LRsymmetry}). To maintain this $Z_2$ symmetry in low-energy effective theory, the UV completion should be invariant under colored preons' LR exchange symmetry
\bea
\chi_L \leftrightarrow \chi_R.
\eea
The flavor symmetry in colored preon sector is $SU(6)\times U(1)$ with respect to hypercolor group with $SU(3)_c \subset SU(6)$.  At low energy scale the hypercolor gauge interaction confine, the colored preons may condense with each other $\langle \chi_i \chi_j \rangle \neq 0$ with $\chi =\{\chi_L,  \chi_R ^c\}$. The colored preons are in
2-index anti-symmetric representation of hypercolor so the condensates are in 2-index symmetric representation of $SU(6)$ global group and thus the global symmetry  would broken into $SO(6)$, resulting in $\bf 20$ colored pNGB. If the symmetric condensates preserve QCD, the quantum number of these pNGBs under $SU(3)_c $ is
\bea
\bf 20 = \bf 8 + 6 +\bar{6}.
\eea
Since these pNGBs are colored, they can be produced through QCD with big cross section at LHC, resulting in around $1$ TeV lower bounds for their masses~\cite{Cacciapaglia:2015eqa}. However, their masses can be from several sources, such as QCD interactions, the mass of colored preons and fermion Yukawas. Combining these effects, its mass can be heavy enough to escape experiment searches. More study on this UV completion can be found in~\cite{Csaki:2017jby}.

\section{Higgs potential and Fine Tuning}\label{sec:tuning}
In this section we fully calculate the Higgs potential and discuss the EWSB. According to eq.(\ref{Vf}), the contribution of fermion loops to Higgs potential could be parameterized as follows (up to $\mathcal{O}(s_h^4)$ order),
\begin{equation}\label{Vfermion}
  V_f = -\gamma_f s_h^2 + \beta_f s_h^4,
\end{equation}
while the contribution from gauge loops is~\cite{Csaki:2017jby}
\begin{equation}
  V_g = \frac{3f^2(3g^2+g'^2-2g_1^2)m_\rho^2\ln 2}{64\pi^2}s_h^2 \equiv \gamma_g s_h^2.
\end{equation}
The gauge contribution to $s_h^4$ can be neglected since it is at $\mathcal{O}(g^4)$ order. So the total Higgs potential has the form as
\begin{equation}\label{eq:GenPotential}
  V = -\gamma s_h^2 + \beta s_h^4,
\end{equation}
where $\gamma=\gamma_f-\gamma_g$ and $\beta=\beta_f$. The overall $\gamma$ has to be positive for a realistic model and the potential has a minimum at $s_h^2=\frac{\gamma}{2\beta}\equiv\xi$. The exact $Z_2$ symmetry in the fermion sector leads to $\gamma_f=\beta_f$ for the leading expression so the natural $\xi$ from the fermion sector is about $\xi\approx 0.5$. In order to get an experimentally allowed small $\xi$ ($\xi\ll1$) to realize correct EWSB, the fermionic and gauge contribution to $\gamma$ should have a cancellation which brings some fine tuning to this model. For the realistic VEV $\xi\ll1$, the top mass can be extracted from eq.(\ref{eq:top_effective}) as
\begin{equation}\label{eq:top_mass}
  m_t=\left.\frac{|M^t_1(p)|s_h}{\sqrt{2\Pi_0(p)(2\Pi_0(p)+\Pi_1(p))}}\right|_{p\rightarrow0}=y_tv,
\end{equation}
where $\Pi^q_{0,1}=\Pi^t_{0,1}\equiv\Pi_{0,1}$, $y_t$ and $v$ are the SM top Yukawa coupling and Higgs VEV. From the explicit expressions of form factors in eq.(\ref{eq:Form Factor}), one can find $M^t_1(p)\simeq M_f \Pi_1(p)$ where $M_f$ is a typical top partner mass. So use equation (\ref{eq:top_mass}), the leading expression of $\gamma_f$ and $\beta_f$ (up to $\mathcal{O}(y_t^2)$) can be estimated from eq.(\ref{Vf}) by power counting as
\begin{equation}
  \gamma_f=\beta_f\simeq\frac{4N_c}{(4\pi)^2}y_t^2f^2M_f^2.
\end{equation}
The Higgs mass then can be derived from the potential (\ref{eq:GenPotential}) as
\begin{equation}
 m_h^2 = \frac{8\beta\xi(1-\xi)}{f^2}\simeq\frac{2N_c\xi(1-\xi)y_t^2}{\pi^2}M_f^2.
\end{equation}
We see that the Higgs mass is sensitive to the top partner mass in this model, so the top partner can't be too heavy for a light Higgs. If we fix $\xi=0.03$, $m_h=125$ GeV and $m_t=150$ GeV, the top partner mass is about $M_f\simeq1.54$ TeV.

The tuning can be quantified as~\cite{Panico:2012uw}
\begin{equation}
 \Delta=\text{Max}(\Delta_i) ,\quad \Delta_i=\left|\frac{\partial\ln\xi}{\partial\ln x_i}\right| ,
\end{equation}
where $x_i$ are the free parameters in the model. Since the $Z_2$ symmetry has fixed $\gamma_f\approx\beta_f$ the main tuning is from the cancellation between $\gamma_f$ and $\gamma_g$. Moreover, $M_f$ is almost fixed by Higgs mass thus the main tuning comes from the variation of free parameters $m_\rho$ and $U(1)_\eta$ gauge coupling $g_1$. The explicit expression for maximal tuning $\Delta$ is
\begin{align}\label{delta1}
  \Delta  =\Delta_{m_\rho}&=\frac{1}{\xi}-2 \: \text{for}\: m_\rho<\frac{4\pi m_h\sqrt{1-2\xi}}{\sqrt{3\xi(1-\xi)(3g^2+g'^2)\ln2}} \nonumber\\
  \Delta  =\Delta_{g_1}&=\frac{1-2\xi}{\xi}\left(\frac{3\xi(1-\xi)(3g^2+g'^2)m_\rho^2\ln2}{8\pi^2m_h^2(1-2\xi)}-1\right) \text{for}\nonumber\\
  m_\rho &> \frac{4\pi m_h\sqrt{1-2\xi}}{\sqrt{3\xi(1-\xi)(3g^2+g'^2)\ln2}}.
\end{align}
For $\xi=0.03$, the tuning is about $\Delta\simeq30$ if the gauge resonance mass $m_\rho$ less than 5 TeV. When $m_\rho>5$ TeV, the tuning will grow linearly with $m_\rho^2$ because the cancellation between $U(1)_\eta$ and SM gauge sector becomes significant and will dominate the tuning.

\begin{figure}
\begin{center}
\includegraphics[width=0.49\columnwidth]{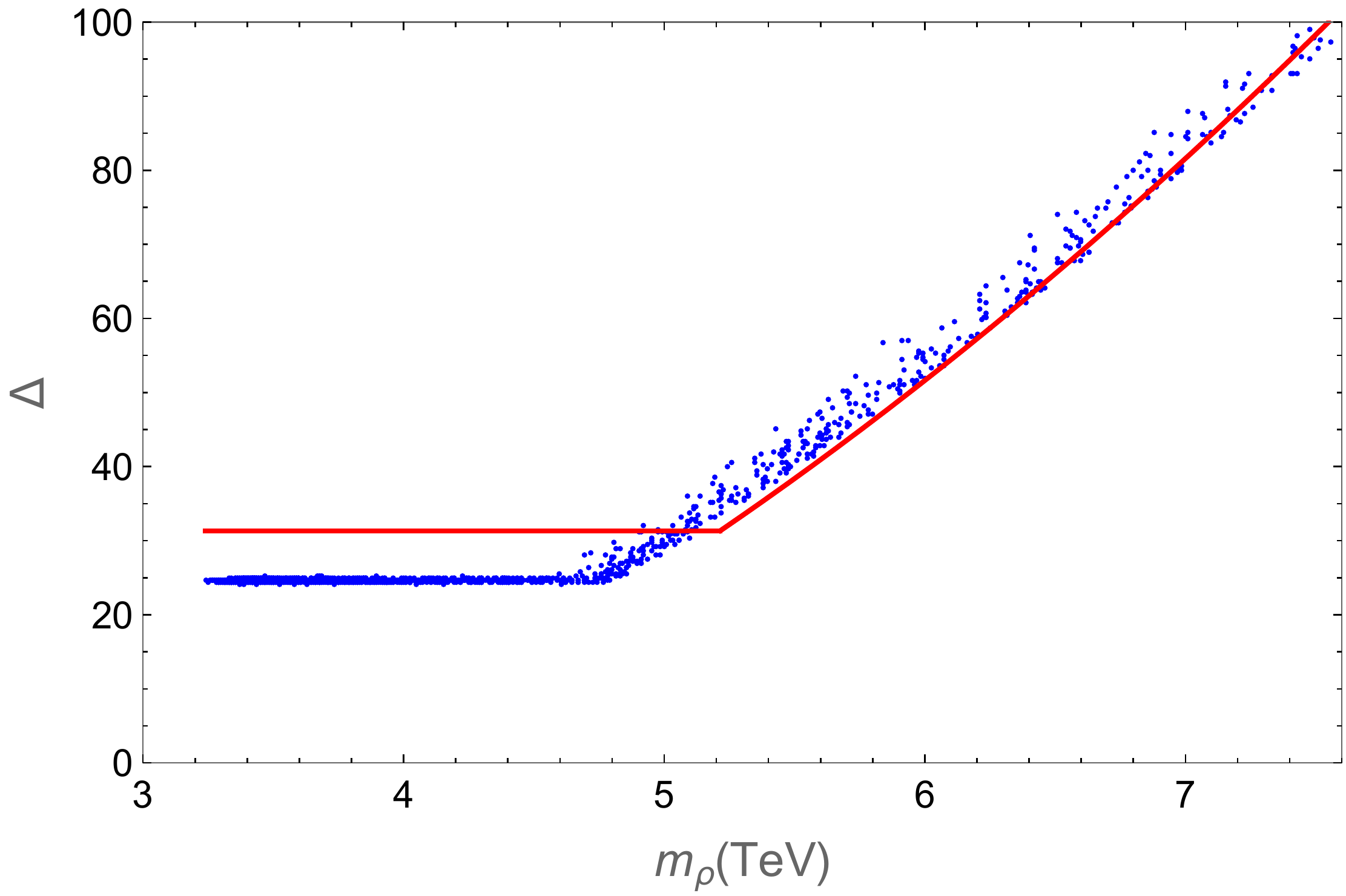}
\includegraphics[width=0.49\columnwidth]{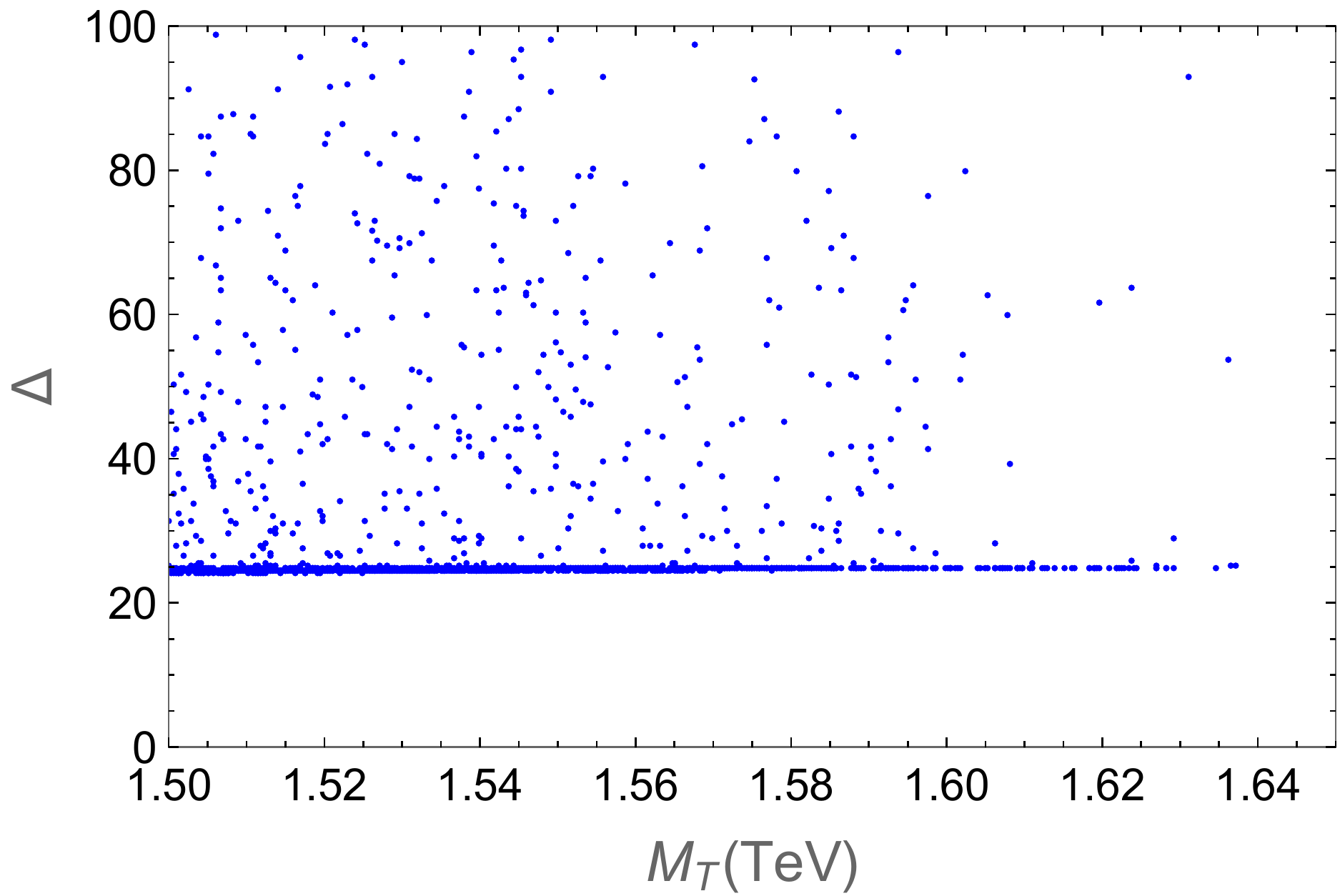}
\end{center}
\caption{Scatter plot for $\xi=0.03$, $m_t\in[140,170]$ GeV, $m_h\in[120,130]$ GeV and $U(1)_\eta$ gauge boson mass $m_{B\prime}\in[\frac{125}{2},1000]$ GeV. In the left (right) panel we show the tuning $\Delta$ as a function of $m_\rho$ (lightest top partner $M_T$). The red line corresponds to the estimation from eq.(\ref{delta1}).}\label{grho}
\end{figure}
\begin{figure}
  \centering
  \includegraphics[width=8cm]{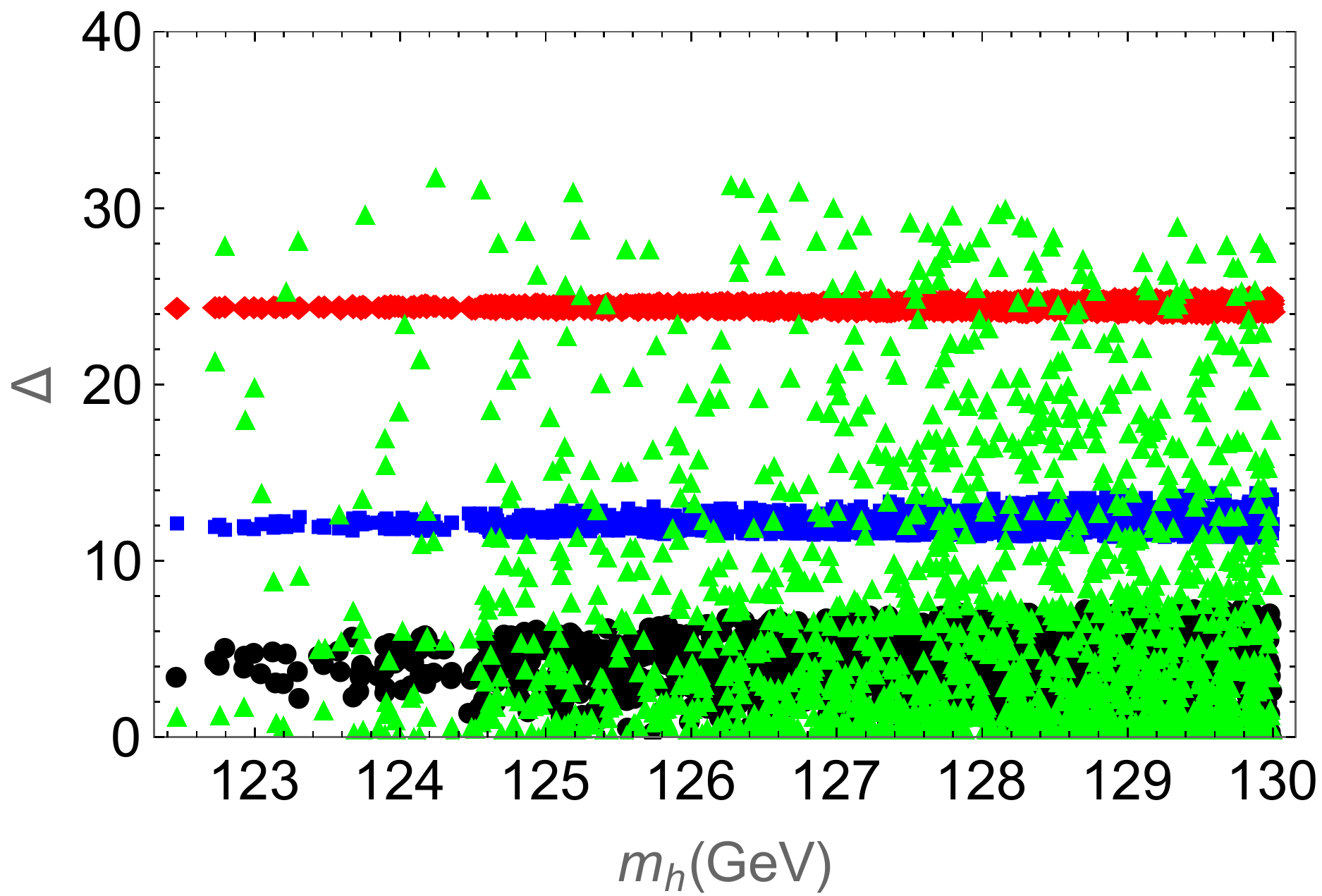}\\
  \caption{Scatter plot of tuning $\Delta_i$ as a function of $m_h$ with $\xi$ fixed at 0.03 for the various free input parameters $x_i$: $M_5$(black),$M_1$(blue), $m_\rho$(red) and $g_1$(green). We choose the range of parameters as follows: $m_t\in[140, 170]$ GeV, $m_\rho\in[2.5,5]$ TeV, $m_{B\prime}\in[\frac{125}{2},1000]$ GeV and the lightest top partner's mass $M_T>1.5$ TeV}\label{tunemh}
\end{figure}

In Fig.\ref{grho}, we show the exact numerical result of tuning $\Delta$ as a function of $m_\rho$ and $M_T$ for Higgs mass $m_h\in[120,130]$ GeV with $\xi=0.03$, where $M_T$ is the lightest top partner's mass and the red line is the analytical result given by eq.(\ref{delta1}). The deviation between eq.(\ref{delta1}) and the actual result comes from the higher order correction ($\mathcal{O}(y_t^4)$) to $\beta_f$, which finally lead to $\beta_f>\gamma_f$. So the actual tuning of the model is 25 for a light $m_\rho$. The right panel shows that both low tuning $\Delta\sim25$ and heavy top partners can be achieved for a light Higgs. In Fig.\ref{tunemh}, we show the tuning for all the free parameters.

\section{Experimental Signals Related To $U(1)_\eta$ Gauge Boson}\label{sec:phenomonology}

We have never stopped looking for a SM singlet $U(1)$ gauge boson $Z'$ in the experiment, and many BSM models have predicted the existence of such a $Z'$. While in this paper $Z'$ is just the $SO(2)_\eta$ gauge boson $B'$. In this model $B'$ could couple to two fermions through the covariant derivative terms. Since only the right handed top (bottom) and $T^\prime_\pm$ carries the $U(1)_\eta$ charge, the primary decay channel of $B'$ will be $t\bar{t}$ or $b\bar{b}$. In order to avoid the experimental restrictions from Higgs decay channel, we choose the $B'$ mass in the range [125/2,1000] GeV when scanning the parameter space. We simulated the process $pp\rightarrow B'\rightarrow t\bar{t}(b\bar{b})$ and find that the cross section is below 0.31 pb if $m_{B'}<1$ TeV, which is far below the experimental bounds~\cite{ATLAS:ttbar,CDF:bbbar}. So the $B'$ mass we chosen in the model is reasonable and doesn't conflict to the direct search. Notice that the potential six top signals $pp \rightarrow t' \bar{t}' \rightarrow B' t B' \bar{t} \rightarrow ttt \bar{t} \bar{t} \bar{t}$ ($t^\prime$ generally refers to top partners)  will also be the important channel to discover or constrain this model~\cite{Han:2018hcu}.

\section{conclusions}
We introduced a $Z_2$ exchange symmetry between LH and RH fermions to scenario of composite Higgs models which trigger the $s_h\leftrightarrow c_h$ exchange symmetry in the Higgs potential, just like the twin Higgs models. This $Z_2$ is very efficient to reduce tuning because it could directly eliminate the quadratic divergences as well as the so called ``double tuning" in the Higgs potential. If we further impose some additional structure, such as two-site de-construction or enlarge the global symmetry in the mixing terms between elementary and composite sector, the remaining logarithmic divergence could be eliminated and realize a finite Higgs potential with minimal tuning.

The concrete model we constructed is based on the symmetric coset $SO(6)/SO(5)\simeq SU(4)/Sp(4)$ so there are 5 NGB fields which include a Higgs doublet and a SM singlet $\eta$. We gauge the SM groups as well as $U(1)_\eta$ so $\eta$ will be eaten by the $U(1)_\eta$ gauge boson $B'$ after $U(1)_\eta$ is spontaneously broken by strong dynamics. One advantage of this model is that the isomorphic $SU(4)/Sp(4)$ coset has very well defined pure fermionic UV completions. Although the gauge sector can't realize such LR symmetry, the fermionic UV can force the composite resonances to satisfy Weinberg sum rules and thus Higgs potential is converge. The successful EWSB ($\xi \ll 1$) is realized by imposing the cancellation between top and gauge sector, which is the main source of tuning in this model. The numerical calculation shows that a light Higgs can be realized with heavy top partners ($>1.5$ TeV) and gauge resonances ($>3$ TeV), which brings a 4\% tuning for $\xi=0.03$.

\section*{Acknowledgements}
T.M. is supported in part by project Y6Y2581B11 supported by 2016 National Postdoctoral Program for Innovative Talents in China and also supported by the Israel Science Foundation (Grant No. 751/19), the United States-Israel Binational Science Foundation (BSF) (NSF-BSF program Grant No. 2018683) and the Azrieli foundation. J.S. is supported by the National Natural Science Foundation of China (NSFC) under grant No.11847612, No.11690022, No.11851302, No.11675243 and No.11761141011, and also supported by the Strategic Priority Research Program of the Chinese Academy of Sciences under grant No.XDB21010200 and No.XDB23000000.

\appendix

\section{Gauge Symmetry In Elementary Sector}
\label{app:gauge}
The coset space $SO(6)/SO(5)$ contains $5$ NGBs parametrized by $h_i$ and $\eta$ with $i=1,2,3,4$. The custodial symmetry $SO(4) \cong SU(2)_L \times  SU(2)_R$ is embedded in the first four indices of $SO(6)$. Under the custodial symmetry, $h_i$ and $\eta$ is $SO(4)$ quartet and singlet so $h_i$ corresponds to the SM Higgs doublet. These NGBs can be described by non-linear sigma field
\bea
  U=\mbox{ exp} ( i\sqrt{2}\pi^{\hat{a}} T^{\hat{a}}/f ),
\eea
where $\pi^{\hat{a}}=\{h_i ,\eta \}$ and $T^{\hat{a}}$ are broken generators associated with the NGBs and normalized as $\mbox{Tr}[T^{\hat{a}} T^{\hat{b}}] =\delta^{\hat{a} \hat{b}}$. The $SU(2)_L \times U(1)_Y $ with $U(1)_Y \subset SU(2)_R$ is gauged to be EW symmetry. Besides, we also gauge an $SO(2)_\eta$ which is the rotation between the last two indices of $SO(6)$. Since $SO(2)_\eta$ is in the broken direction, the associated NGB $\eta$ will be eaten by the corresponding massive gauge boson. So after EW symmetry is broken by vacuum misalignment, only one pNGB remains to play the role of Higgs and is denoted by $h$.

The gauge interactions of the pNGBs can be written in terms of linear realized $\Sigma$ field and the leading order Lagrangian is given by
\bea
\mathcal{L}=\frac{f^2}{2} (D_\mu \Sigma )^T D^\mu \Sigma,
\eea
where $D_\mu = \partial_\mu -i g W_\mu ^a T^a_L -i g^\prime B_\mu T_R ^3 -i g_1 B_\mu ^\prime T_\eta $ and $T_L ^a$, $T_R ^3$ and $T_\eta$ is generators of $SU(2)_L$, $U(1)_Y$ and $SO(2)_\eta$ embedded in $SO(6)$. After EW is broken, the gauge bosons masses are
\bea \label{eq:boson_mass}
m_W =\frac{g f}{2} s_h,\quad m_Z =\frac{m_W}{\cos \theta_W},\quad m_{B^\prime} = \frac{g _1 f}{\sqrt{2}}c_h,
\eea
where $\theta_W$ is weak mixing angle.

\section{Generators}
\label{App:generators}
In this appendix we list the explicit expressions of $SO(6)$ and $SU(4)$ generators we used in our model.

\subsection{$SO(6)/SO(5)$}
The $SO(6)$ generators could be classified in five broken generators in $SO(6)/SO(5)$ coset, six unbroken generators of custodial symmetry $SO(4)\cong SU(2)_L\times SU(2)_R$ and four unbroken generators of $SO(5)/SO(4)$ coset
\begin{align}\label{App:SO_6_generator}
  T^{\hat{a}}_{ij} & =-\frac{i}{\sqrt{2}}(\delta^{\hat{a}i}\delta^{6 j}-\delta^{\hat{a}j}\delta^{6 i}),\nonumber \\
  T^a_{L,Rij} & =-\frac{i}{2}[\epsilon^{abc}\delta^{bi}\delta^{cj}\pm(\delta^{ai}\delta^{4 j}-\delta^{aj}\delta^{4 i})],\nonumber\\
  T^{\alpha}_{ij} & =-\frac{i}{\sqrt{2}}(\delta^{\alpha i}\delta^{5 j}-\delta^{\alpha j}\delta^{5 i}),
\end{align}
where $\hat{a}$ from 1 to 5, $a$ from 1 to 3 and $\alpha$ from 1 to 4. The generator of $SO(2)_\eta$ is $T^{\hat{5}}\equiv T^{\eta}$.

\subsection{$SU(4)/Sp(4)$}
The global symmetry breaking VEV $V$ in eq.(\ref{eq:vacuum}) breaks the global $SU(4)$ to $Sp(4)$. The unbroken and broken generators satisfy the identity as follows:
\begin{equation}
  T^{\alpha}=-V T^{\alpha T}V^T, \quad T^{\hat{\alpha}}=V T^{\hat{\alpha}T}V^T.
\end{equation}
So the explicit generators could be chosen as follows. The generators of custodial symmetry $SU(2)_L\times SU(2)_R$ are:
\begin{equation}
  T^a_L=\frac{1}{2}\left(
                     \begin{array}{cc}
                       \sigma^a & 0 \\
                       0 & 0 \\
                     \end{array}
                   \right) ,\quad T^a_R=\frac{1}{2}\left(
                                                     \begin{array}{cc}
                                                       0 & 0 \\
                                                       0 & -\sigma^{aT} \\
                                                     \end{array}
                                                   \right).
\end{equation}
The remaining four unbroken generators are:
\begin{equation}
  \frac{1}{2\sqrt{2}}\left(
                       \begin{array}{cc}
                         0 & i\sigma^a \\
                         -i\sigma^a & 0 \\
                       \end{array}
                     \right) ,\quad \frac{1}{2\sqrt{2}}\left(
                                                         \begin{array}{cc}
                                                           0 & \mathds{1}_2 \\
                                                           \mathds{1}_2 & 0 \\
                                                         \end{array}
                                                       \right).
\end{equation}
The five broken generators are:
\begin{align}
  T^{\hat{a}} & =\frac{1}{2\sqrt{2}}\left(
                                      \begin{array}{cc}
                                        0 & \sigma^a \\
                                        \sigma^a & 0 \\
                                      \end{array}
                                    \right) , T^{\hat{4}}=\frac{1}{2\sqrt{2}}\left(
                                                                               \begin{array}{cc}
                                                                                 0 & i\mathds{1}_2 \\
                                                                                 -i\mathds{1}_2 & 0 \\
                                                                               \end{array}
                                                                             \right),
   \nonumber\\
  T^\eta & =\frac{1}{2\sqrt{2}}\left(
                                 \begin{array}{cc}
                                   \mathds{1}_2 & 0 \\
                                   0 & -\mathds{1}_2 \\
                                 \end{array}
                               \right),
\end{align}
where as in $SO(6)/SO(5)$ case, $T^\eta$ is the generator of $U(1)_\eta$.

\section{Form Factors In Effective Lagrangian}
\label{App:form factor}
We can get the form factors in effective Lagrangian eq.(\ref{eq:top_effective}) by integrating out the fermion resonances from the global $SO(6)$ invariant Lagrangian eq.(\ref{eq:Lagrangian}). The explicit expression is:
\begin{align}\label{eq:Form Factor}
  \Pi^q_0(p) & =1+\frac{f^2\epsilon_{q5}^2}{M_5^2-p^2},\;\Pi^q_1(p)  =\frac{f^2\epsilon_{q1}^2}{M_1^2-p^2}-\frac{f^2\epsilon_{q5}^2}{M_5^2-p^2}\nonumber \\
  \Pi^t_0(p) & =1+\frac{f^2\epsilon_{t5}^2}{M_5^2-p^2},\;\Pi^t_1(p) =\frac{f^2\epsilon_{t1}^2}{M_1^2-p^2}-\frac{f^2\epsilon_{t5}^2}{M_5^2-p^2} \nonumber\\
  M^t_1(p) & =f^2\left(\frac{M_1\epsilon_{q1}\epsilon_{t1}}{M_1^2-p^2}-\frac{M_5\epsilon_{q5}\epsilon_{t5}}{M_5^2-p^2}\right).
\end{align}

\section{Some comments on bottom sector}
Since the the global symmetry $SO(6)$ is not big enough to put the RH bottom and top in the same multiplet (see Eq.(\ref{eq:embedding})), actually LR exchange symmetry in kinetic terms is broken if include both the bottom and top sector interactions.  However as discussed in~\cite{Csaki:2017jby}, if the bottom is assigned in a proper representation the $Z_2$ symmetry can be broken collectively such that the leading Higgs contributions from broken sector are at order $\mathcal{O}(\epsilon_t ^2 \epsilon_b ^2)$, where $\epsilon_{t,b} $ are characteristic Yukawa coupling in top  and bottom sectors. According to power counting, it is easy to find Higgs potential is still log divergent. To satisfy above requirements, we find the LH bottom doublet and RH bottom singlet should be put in the same representation as top sector
\bea
\Psi_{q_L} ^\prime =\frac{1}{\sqrt{2}}  \left( \begin{array}{c}
t_L \\
-i t_L \\
b_L \\
i b_L \\
0\\
0
\end{array} \right), \quad \Psi_{b_{R}} =  \frac{1}{\sqrt{2}} \left( \begin{array}{c}
0 \\
0 \\
0 \\
0\\
b_R\\
 ib_R
\end{array} \right).
\eea
Since the bottom Yukawa couplings are much smaller than top Yukawas $\epsilon_b \ll \epsilon_t$, the main contribution to Higgs potential is always from $Z_2$ preserving top sector and the $Z_2$ broken contributions, at $\mathcal{O}(\epsilon_t ^2 \epsilon_b ^2)$, can be neglected. So in this paper we only focused on top sector.

\end{document}